\documentclass{article}

\usepackage{arxiv}

\usepackage[utf8]{inputenc} % allow utf-8 input
\usepackage[T1]{fontenc}    % use 8-bit T1 fonts
\usepackage{hyperref}       % hyperlinks
\usepackage{url}            % simple URL typesetting
\usepackage{booktabs}       % professional-quality tables
\usepackage{amsfonts}       % blackboard math symbols
\usepackage{nicefrac}       % compact symbols for 1/2, etc.
\usepackage{microtype}      % microtypography
\usepackage{lipsum}
\usepackage{graphicx}
\graphicspath{ {./images/} }
\usepackage{changepage}
\usepackage{epsfig}
\usepackage{epstopdf}
\usepackage{xcolor}
\usepackage{textcomp}
\usepackage{manyfoot}
\usepackage{algorithm}
\usepackage{algorithmicx}
\usepackage{algpseudocode}
\usepackage{listings}
\usepackage{caption}
\usepackage{subcaption}
\usepackage[numbers]{natbib}
\usepackage{pifont}
\usepackage{amsmath}

\title{Future Industrial Applications: Exploring LPWAN-Driven IoT Protocols}

\author{
 Mahbubul Islam \\
  Department of Computer Science\\
  United International University\\
  Dhaka, Bangladesh \\
  \And
   Hossain Md. Mubashshir Jamil \\
  Department of Electrical and Electronic Engineering\\
  Islamic University of Technology\\
  Dhaka, Bangladesh \\
  \And
   Samiul Ahsan Pranto \\
  Department of Electrical and Electronic Engineering\\
  Islamic University of Technology\\
  Dhaka, Bangladesh \\
  \And
 Rupak Kumar Das \\
  College Of Information Sciences and Technology\\
  Pennsylvania State University--University Park\\
  State College, PA \\
    \And
 Al Amin \\
  Department of Information Systems\\
  University of Maryland - Baltimore\\
  Baltimore, MD \\
    \And
 Arshia Khan \\
  Department of Computer Science\\
  University of Minnesota - Duluth\\
  Duluth, MN \\
}

\begin{document}
\maketitle
\begin{abstract}
The Internet of Things (IoT) will bring about the next industrial revolution in Industry 4.0. The communication aspect of IoT devices is one of the most critical factors in choosing the suitable device for the suitable usage. So far, the IoT physical layer communication challenges have been met with various communications protocols that provide varying strengths and weaknesses. Moreover, most of them are wireless protocols due to the sheer number of device requirements for IoT. This paper summarizes the network architectures of some of the most popular IoT wireless communications protocols. It also presents a comparative analysis of critical features, including power consumption, coverage, data rate, security, cost, and Quality of Service (QoS). This comparative study shows that Low Power Wide Area Network (LPWAN) based IoT protocols (LoRa, Sigfox, NB-IoT, LTE-M ) are more suitable for future industrial applications because of their energy efficiency,  high coverage, and cost efficiency. In addition, the study also presents an industrial Internet of Things (IIoT) application perspective on the suitability of LPWAN protocols in a particular scenario and addresses some open issues that need to be researched. Thus, this study can assist in deciding the most suitable protocol for an industrial and production field.
\end{abstract}

% keywords can be removed
%\keywords{First keyword \and Second keyword \and More}

\section{Introduction}

\begin{figure}[htbp]
\centerline{\includegraphics[scale=.7]{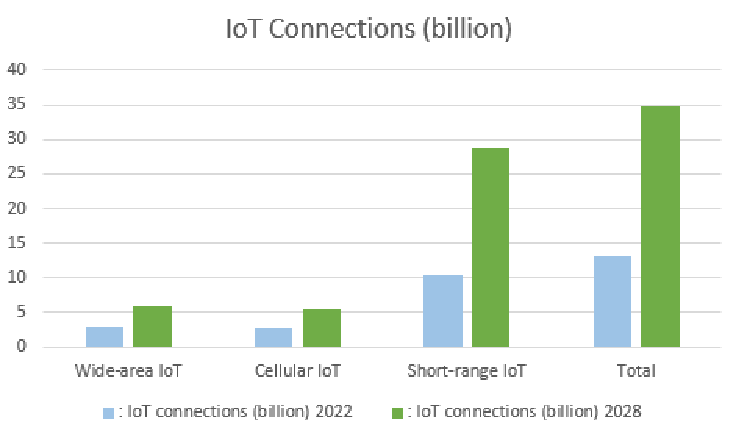}}
\caption{IoT connectivity number forecast}
\label{fig_1}
\end{figure}

Due to the recent emergence of 5G in the wireless telecommunications domain, the Internet of Things (IoT) is taking flight in many aspects of our day-to-day lives. IoT devices need to be able to work at long-ranges and during locomotion. Indeed, connecting numerous such devices under the same network is easier when done on a wireless medium. By 2028, IoT connections have the possibility to increase by around 21.5 billion, with a comparison shown between the number of devices in 2022 and a forecast for 2028 in Figure \ref{fig_1}~\cite{Carson2022-iq}. However, the IoT deployment environment has some different requirements compared to other wireless environments. IoT will mostly be used in smart parking and vehicle-to-vehicle communication, augmented maps, data collection, smart water supply, and smart appliances at homes and offices. Smart power grids, agriculture, and health sectors are also going to be reliant on the interconnectivity of IoT devices. The deployment of IoT devices has some requirements that differ from conventional wireless telecommunication requirements. For example, a huge factor of IoT systems is scalability from a number of devices' points of view. One of its characteristics is the massive number of devices it incorporates in a network, and it performs upon frugal power consumption by inexpensive devices.
According to Figure \ref{fig_1}, we have found that current trends and future trends both show that wide area protocols and short-range protocols supported devices are mainly used while cellular protocols supported devices are comparatively less in use. Considerable research(cite) supported the idea that broad areas and short ranges, primarily LPWAN-based protocols, are superior to cellular IoT protocols because of their performance.
Three of the key requirements of IoT technologies are low power consumption and long-range communication at low cost. Besides these, parameters like data rate, security, link budget, and others should also be taken into consideration. Different technologies occupying IoT telecommunications fulfill these requirements differently. Some of them offer a better range, while others offer lower power consumption. In order to decide which technology should be the most suitable in a certain deployment environment, it is essential to put these parameters side by side and get a clear view. Some of the most popular IoT communication protocols used today are Z-Wave, LoRa, the Narrow Band Internet of Things (NB-IoT), Sigfox, and Long Term Evolution for Machines (LTE-M).

On one hand, there are IoT protocols like Lora, Sigfox, and Z-Wave that operate in the unlicensed ISM spectrum. LoRa is a deep-rooted exclusive framework in the IoT business that utilizes the Macintosh-level convention LoRaWAN and offers long reach and an extraordinary geo-inclusion~\cite{augustin2016study}.  Sigfox is one of the least power-consuming IoT frameworks right now~\cite{gomez2019sigfox}. On the other hand, systems like NB-IoT and LTE-M make use of the licensed frequency spectrum, resulting in higher power consumption but better quality. NB-IoT is a high-quality IoT technology that operates on a licensed spectrum and was made available in the 3rd Generation Partnership Project (3GPP) Release 13. It can support a large number of devices in a single cell, has a larger coverage area, has low device complexity, and offers flexibility in deployment~\cite{wang2017primer}. Another LTE-based protocol for machine-to-machine (M2M) communication, LTE-M, was also included in 3GPP Release 13 and operates at the licensed frequency bands~\cite{ratasuk2017lte}. 

There have been numerous approaches in order to present a comparison among the different IoT protocols. Some surveys present a comparison focusing on various IoT-based wireless sensor networks (WSN) but lack real-time testing~\cite{aldahdouh2019survey}. Studies in~\cite{gloria2017comparison} and~\cite{al2017internet} show a comparison among the devices regarding various parameters, albeit in a limited environment. All these surveys focus on various metrics and face distinct challenges, but to our best knowledge, not much attention has been given to three essential comparison matrices simultaneously, power, range, and cost, and the identification of the most suitable IoT protocol for future industrial use.

In this study, we compare and contrast the most widely used Internet of Things deployment standards in terms of physical layer communication protocols like Sigfox, LoRa, Z-Wave, NB-IoT, and LTE-M and their parameters such as power utilization, cost, range, data rate, QoS, and security. In order to comprehend the fundamental differences between the technologies, we also provide a summary of their network architectures. In order to select the appropriate protocol for a given application scenario, we also examine the five protocols from an application perspective. Comparative scrutiny has shown that LPWAN-based protocols are the most suitable for upcoming internet applications due to their optimal capacity, cost-effectiveness, and energy efficiency. We have therefore demonstrated that LPWAN protocols are the protocols of the future for the upcoming industrial revolution and have offered the optimal application perspective for the industrial Internet of Things (IoT). Finally, we discuss some of the key difficulties present in the current writing in this new area of examination that require attention and are open for investigation in future exploration works regarding IoT correspondence.

The rest of the article is illustrated as follows: Section \ref{section_Literature} provides a brief summary and contribution to a comparative analysis of existing literature. Section \ref{section_protocols} describes available wireless IoT protocols. Section \ref{section_performance} contains a performance comparison of the targeted IoT protocols. Section \ref{section_perspective} discusses the Industrial application perspective of the relevant IoT protocols and summarizes them. Section \ref{section_future} involves the scope for future works described in the surveyed literature. Finally, section \ref{section_conclusion} brings a conclusion to the survey carried out in this research.

%%%%%%%%%%%%%%%%%%%%%%%%%%%%%%%%%%%%%%%%%%
\section{Literature Survey}\label{section_Literature}

In this section, we have provided a brief overview and synthesis of significant research. Most of the research is based on comparative analysis among the IoT physical layer protocols based on technology, application, and performance.

\begin{table*}[!htb]
\centering
%\begin{adjustwidth}{-2cm}{-2cm}
\scriptsize
\caption{Comparative analysis of existing literature}
\label{table:Related work}
%\begin{tabular}{|m{2em}|m{2cm}|m{4.5cm}|m{4cm}|m{1.55cm}|m{1.55cm}|}
\begin{tabular}{cccccc}
\hline
\textbf{Research}                                                 & \textbf{IoT Layer}                                             & \textbf{Metrics}                                                                                                             & \textbf{Protocols}                           

& \textbf{\begin{tabular}[c]{@{}c@{}}Comparative \\ Analysis\end{tabular}} & \textbf{\begin{tabular}[c]{@{}c@{}}Experimental  \\ Analysis\end{tabular}} \\ \hline \\
\begin{tabular}[c]{@{}c@{}}Andre  Gl´oria \\ et al~\cite{gloria2017comparison}\end{tabular} & \begin{tabular}[c]{@{}c@{}}Physical, \\ Data link\end{tabular} & \begin{tabular}[c]{@{}c@{}}Multinode capability, low-cost and \\ power saving capabilities, \\ delay, data rate\end{tabular} & \begin{tabular}[c]{@{}c@{}}Wifi, Zigbee,  LoRa, \\ Bluetooth\end{tabular}                        & \checkmark                                                                      & \checkmark                                                                        \\  \\ %\hline
\begin{tabular}[c]{@{}c@{}}Shadi  Al-Sarawi\\  et al~\cite{al2017internet}\end{tabular} & \begin{tabular}[c]{@{}c@{}}Physical, \\ Data link\end{tabular} & \begin{tabular}[c]{@{}c@{}}Power consumption, security, \\ spreading data rate\end{tabular}                                  & \begin{tabular}[c]{@{}c@{}}6LoWPAN, ZigBee, BLE, \\ Z-Wave , NFC , SigFox  \\ \\ LPWAN\end{tabular} & \checkmark                                                                      & \checkmark                                                                        \\ %\hline
\begin{tabular}[c]{@{}c@{}}Ala’  Khalifeh \\ et al~\cite{aldahdouh2019survey}\end{tabular}   & \begin{tabular}[c]{@{}c@{}}Physical, \\ Data link\end{tabular} & \begin{tabular}[c]{@{}c@{}}Small size, low cost, \\ limited energy\end{tabular}                                              & \begin{tabular}[c]{@{}c@{}}LoRaWAN, Sigfox,  NB-IoT \\ and LTE-M\end{tabular}                    & \checkmark                                                                      & \ding{55}                                                                      \\\\ %\hline
\begin{tabular}[c]{@{}c@{}}Burak  H. Çorak \\ et al~\cite{ccorak2018comparative}\end{tabular}  & Application                                                    & \begin{tabular}[c]{@{}c@{}}Packet creation time, packet delivery\\  speed, delay differences\end{tabular}                    & CoAP, MQTT and XMPP                                                                              & \checkmark                                                                     & \checkmark                                                                         \\\\ %\hline
\begin{tabular}[c]{@{}c@{}}Thays  Moraes \\ et al~\cite{moraes2019performance}\end{tabular}    & Application                                                    & \begin{tabular}[c]{@{}c@{}}Throughput, message size,\\ packet loss\end{tabular}                                              & AMQP, CoAP and MQTT                                                                              & \checkmark                                                                      & \checkmark                                                                        \\\\ %\hline
JASENKA  et al~\cite{dizdarevic2019survey}                                                    & Application                                                    & \begin{tabular}[c]{@{}c@{}}Latency, energy consumption and \\ network throughput\end{tabular}                                & \begin{tabular}[c]{@{}c@{}}MQTT, AMQP, XMPP,\\ DDS, HTTP and CoAP\end{tabular}                & \checkmark                                                                      & \ding{55}                                                                         \\\\ %\hline
\begin{tabular}[c]{@{}c@{}}Kais  Mekki \\ et al~\cite{mekki2018overview}\end{tabular}      & Physical                                                       & \begin{tabular}[c]{@{}c@{}}Network capacity, devices lifetime, \\ cost, quality of service and latency\end{tabular}          & Sigfox, LoRaWAN and NB-IoT                                                                       & \checkmark                                                                      & \ding{55}                                                                          \\\\ %\hline
H. Mroue  et al~\cite{mroue2018mac}                                                   & Physical                                                       & \begin{tabular}[c]{@{}c@{}}Carrier frequency, packet duration, \\ number of channels \\ and spectrum access.\end{tabular}       & LoRa, Sigfox and NB-IoT                                                                          & \checkmark                                                                      & \checkmark                                                                         \\\\ %\hline
Our Research                                                      & \begin{tabular}[c]{@{}c@{}}Physical, \\ Data link\end{tabular}                                                     & \begin{tabular}[c]{@{}c@{}}QoS, security, \\ power consumption, cost, \\ coverage, datarate\end{tabular}                               & \begin{tabular}[c]{@{}c@{}}LoRa, Sigfox , NB-IoT\\ LTE-M, Z-Wave\end{tabular}                    & \checkmark                                                                      & \ding{55}                                                                          \\ \hline
\end{tabular}
%\end{adjustwidth}
\end{table*}

In order to determine the optimal communication protocol, Andre Gl’oria et al. thoroughly study the key protocols currently in use, conduct a comparison analysis, and then select protocols based on the findings. According to their findings, LoRa is a more trustworthy option for a wireless protocol since it requires little complexity and expense~\cite{gloria2017comparison}. Shadi Al-Sarawi et al. overview IoT communication protocol visions, advantages and disadvantages, and additional QoS like energy consumption range and data rate to compare IoT communication protocols~\cite{al2017internet}.
  
Ala’ Khalifeh et al. evaluate several wireless technologies (LoRaWAN, NB-IoT, Sigfox, and LTE-M) to consider how they might be employed in fifth-generation (5G) communication technologies and wireless sensor networks~\cite{aldahdouh2019survey}.
According to Thays Moraes et al. studies, throughput, message size, and packet loss were used to evaluate how the Advanced Message Queuing Protocol (AMQP), MQ Telemetry Transport (MQTT), and Constrained Application Protocol (CoAP) behaved concerning speed and fault injection. According to tests, the researchers suggest that the CoAP protocol is an intriguing option for applications with limited network resources~\cite{moraes2019performance}. A concentrate by Mroue et al. describes the characteristics of the Medium Access Control (MAC) layer for the Low Power Wide Area Network (LPWAN) options that are currently available, such as LoRa, Sigfox, and NB-IoT. The displaying of a thick organization for every one of these innovations is likewise canvassed in this review. The characteristics of various systems, such as carrier frequency, packet length, channel count, and spectrum access, are compared in this research contribution~\cite{mroue2018mac}. The scientist makes a model framework with NB-IoT gadgets, an IoT cloud stage, an application server, and a client application to show the benefits of NB-IoT when joined with other LPWA innovations~\cite{chen2017narrowband}.
 
Researchers introduce a concise examination of the superior organization limit, gadget life span, and cost of LoRaWAN and Sigfox. NB-IoT, then again, performs better concerning inactivity and administration quality. Moreover, they examine the different application situations and the innovation that is best in helping future scholastics and business experts~\cite{mekki2018overview}. Mikhaylov and others show that numerous option LoRaWAN geographies portray and demonstrate a wide range of situations considering the actual layer and zeroing in on the issue of organization security~\cite{mikhaylov2016analysis}.
Adelantado et al. illustrate the features and drawbacks of the LoRaWAN protocol, which are explored when a strategy is taken to decide the use case of LoRaWAN technology; furthermore, in which use cases it does not function~\cite{adelantado2017understanding}. In their review, Rashmi et al. analyze and depict the mechanical contrasts between LoRa and NB-IoT with respect to the actual qualities of network engineering. The examination is introduced as a near and engaging review~\cite{sinha2017survey}. Although the model that Hendrik et al. provide is based on the physical layer, it does illustrate some intriguing insights into the decoding performance in LPWAN when there is packet collision~\cite{lieske2016decoding}.

Table \ref{table:Related work} presents a concise summary of the already existing relevant literature and offers a comparative point of view of the nature of the publications and the value they add to this research area. 

Several studies compared a few IoT protocols in an experimental setting, while many IoT protocols have been evaluated in simulated settings without an experimental setup. Additionally, some studies conducted a theoretical analysis of IoT wireless communication protocols. In contrast to earlier efforts, this study suggests a comparative analysis based on architecture, performance evaluation of specific chosen features of different IoT protocols, and future application of most suitable protocols based on different industrial use cases. Range, cost, and energy consumption-based evaluation are also included because those factors will be vital to future industrial IoT applications. In order to advance the field of research, we also highlight a few common challenges for wireless communication protocols in industrial applications.

%%%%%%%%%%%%%%%%%%%%%%%%%%%%%%%%%%%%%%%%%%
\section{Wireless Protocols Architecture}\label{section_protocols}

This section summarizes the characteristics and features of some notable wireless communication protocols that are used for IoT applications where low power and medium to long ranges are basic requirements.

\subsection{LoRaWAN}
LoRaWAN was developed by the LoRa alliance for applications requiring long-range and low-power Wide Area Networks (WAN)~\cite{sethi2017internet}. It enables highly efficient WANs by reducing the necessity of repeaters between nodes, leveraging its very low power consumption. This increases the overall system efficiency by reducing the number of required devices and also increasing the battery life of the nodes. LoraWAN utilizes the Sub-Ghz band~\cite{Sforza2013-aj}. A noteworthy drawback of LoRaWAN is that the data rate is low, thus catering to some specific IoT applications. The ultra-low power consumption also means that only star topology can be implemented, where all the nodes send information to the central gateway, which communicates with LoRa servers ~\cite{samie2016iot, sornin2015lorawan}. LoRa uses the Chirp Spread Spectrum (CSS) modulation scheme~\cite{chaudhari2020lpwan}.

A CSS transmit signal can be shown as:
\begin{align}
x_k\left[n\right]=\sqrt{\frac{E_s}{N}}exp{\left(j\frac{2\pi}{N}kn\right)c\left[n\right]} \label{eq}
\end{align}

where $E_s$ is the signal energy, $c\left[n\right]$ is the discrete time chirp signal with a period of $N$ and $k$ is the data symbol:

\begin{equation}
k=\sum_{i=0}^{SF-1}{2^i\left[b\right]_i}\label{eq}
\end{equation}

and $b$ is the bit-word as such:

\begin{equation}
b\in\left\{0,1\right\}^{SF}\label{eq}
\end{equation}

The data rate of LoRa transmission, $R_b$ depends on the modulation bandwidth $BW$ and the spreading factor $SF$:

\begin{equation}
R_b=SF\ast\frac{BW^{SF}}{2}\label{eq}
\end{equation}

\subsection{Sigfox}
Sigfox is an open-source technology and the first of its kind. Sigfox was designed to focus on an even longer range and lower data rate compared to LoRaWAN, in which it utilizes Ultra Narrow Band (UNB) frequency. It was designed using a variety of low-power IoT devices, including several sensors and M2M applications. It delegates the complex processing to the cloud instead of the individual nodes to reduce resource consumption~\cite{mekki2018overview}. Star topology can be implemented with Sigfox ~\cite{samie2016iot}.

\begin{figure*}[htbp]
\centerline{\includegraphics[scale=.5]{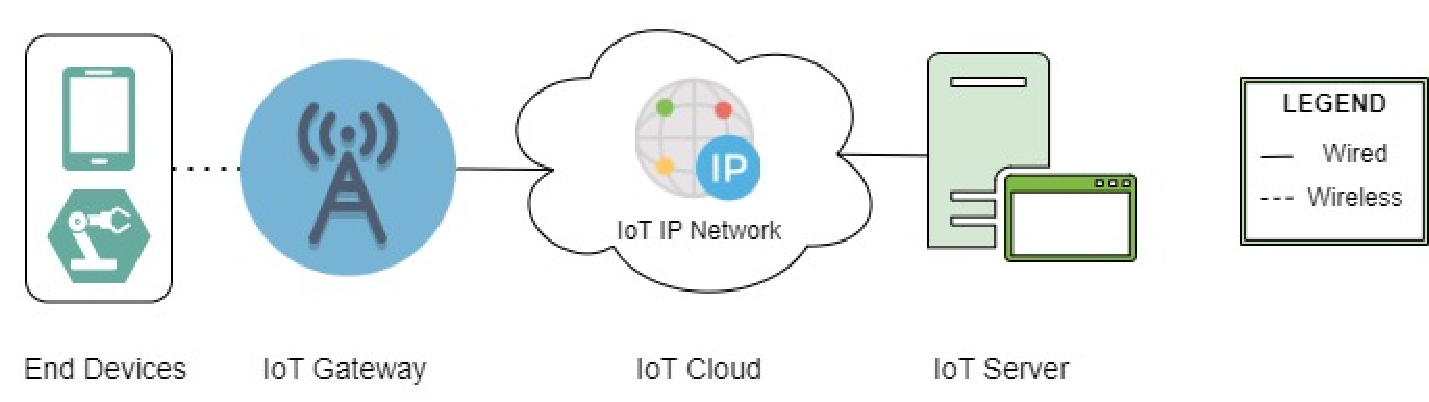}}
\caption{Network architecture for IoT protocols based on unlicensed ISM-band spectrum (Sigfox, LoRa, Z-Wave)}
\label{fig_2}
\end{figure*}

Figure \ref{fig_2} gives a simple overview of the network architecture of IoT technologies that utilize the free-of-cost ISM frequency spectrum to establish and maintain wireless M2M connectivity. In this framework, the data flows from the User End (UE) devices to the IoT base station over the associated wireless protocol: LoRa, Sigfox, or Z-Wave. For example, Sigfox uses its proprietary UNB technologies while utilizing the open-source sub-GHz ISM band spectrum.  
LoRaWAN also employs a sub-GHz ISM band spectrum for wireless connectivity with end devices. The end device can be an actuator, a temperature-measuring device, a sensor, etc. The base station then acts as the gateway to the respective protocol’s IoT server over an ethernet connection. Gateways have point-to-point connectivity with the servers through an IP network, and the data is passed on through the proprietary IoT cloud to the servers. Servers are placed in the network according to requirements, following the specifications supported by the topologies that are being used. Both Sigfox and LoRaWAN support star topology, while Z-Wave can implement mesh topology~\cite{samie2016iot, sornin2015lorawan, fouladi2013security}.

\subsection{NB-IoT}

NB-IoT is an IoT communication protocol that was designed to reduce some of the capabilities of LTE while simultaneously enhancing the ones necessary for IoT networks. This makes NB-IoT an appropriate choice for IoT applications, as it is suitable for low-power IoT devices. As IoT end devices do not need constant back-end broadcasting, the broadcasting frequency is reduced, which decreases the power consumption~\cite{mekki2018overview}. One has to keep in mind that this technology shares the same 3GPP licensed frequency bands with the LTE bands. As a result, only certain operation modes are possible in order to
ensure interference-free operation.

NB-IoT uses the Quadrature Phase Shift Keying (QPSK) modulation scheme to modulate the bits. A QPSK modulated transmit signal can be shown as:

\begin{equation}
S_{QPSK}\left(t\right)=\frac{1}{\sqrt{2T}}a_1\left(t\right)cos{\left(2\pi f_ct+\frac{\pi}{4}\right)}
+\frac{1}{\sqrt{2T}}a_2\left(t\right)sin{\left(2\pi f_ct+\frac{\pi}{4}\right)}\label{eq}
\end{equation}

where a binary bit stream of $a\left(t\right)$ with a period of $T$ and is de-multiplexed into two different bitstreams $a_1\left(t\right)$ and $a_2\left(t\right)$ that represents $+1$ and $-1$ before multiplying by sine and cosine carriers to form a QPSK signal.

\subsection{LTE-M}

LTE-M is another protocol based on the traditional LTE. However, it was designed to transmit bits at a higher data rate at the expense of comparatively higher power utilization and lower battery life compared to NB-IoT. It was designed to cater to applications requiring relatively higher speeds and lower latency. This protocol is useful for certain scenarios where the other protocols do not meet the data-rate requirements. 

\begin{figure*}[htbp]
\centerline{\includegraphics[scale=.5]{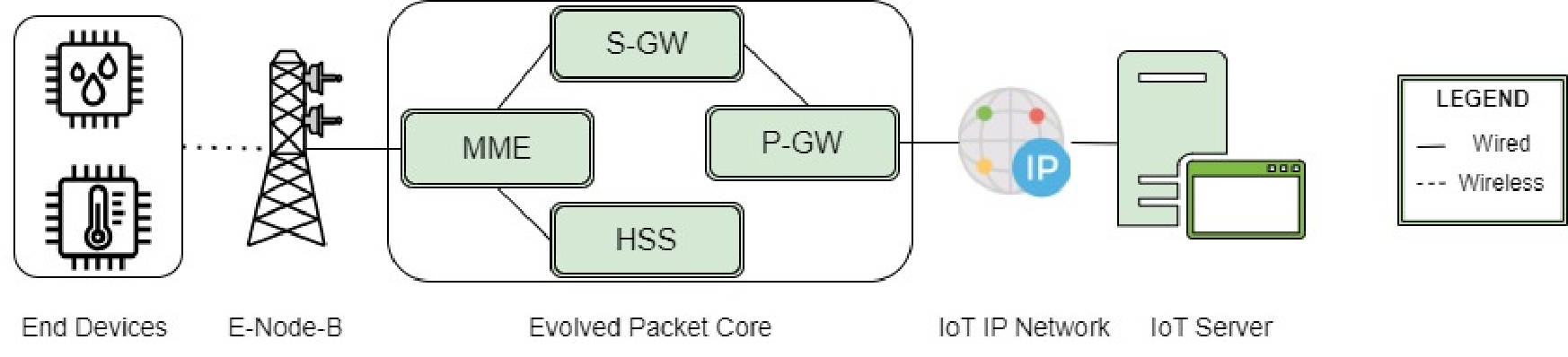}}
\caption{Network architecture for IoT protocols based on 3GPP licensed spectrum (NB-IoT and LTE-M)}
\label{fig_3}
\end{figure*}

Figure \ref{fig_3} illustrates the network connectivity for 3GPP licensed band-based protocols such as NB-IoT and LTE-M. This architecture can be divided into two parts, which are the User Plane and Control Plane. User Plane is concerned with the process of sending and receiving user data. The Control Plane manages the control processes that are implemented to establish communication and authentication of the end devices. The data is transmitted wirelessly from the UE IoT devices to the eNodeB by adopting the LTE-based IoT protocols. The bits are then transferred over a wired network to the Evolved Packet Core (EPC) network. The data is received by the Mobility Management Entity (MME) within the EPC, which focuses on eNodeB signaling, mobility, and security. Home Subscriber Server (HSS) is associated with user authentication and profile. Serving Gateway (SGW) handles the routing and packet forwarding towards the uplink and the downlink. The Packet Gateway (PGW) connects the packet core network with the external IP network. The data then flows through the IP network towards the IoT servers~\cite{zayas20173gpp, iqbal2020application}.

\subsection{Z-Wave}

Z-Wave is a relatively short-range communication protocol widely adopted for IoT applications, which was designed by Sigma Systems~\cite{fouladi2013security, redbend, al2015internet}. It can connect up to 50 devices in a smart home or small-scale commercial environment. Even though it operates in the ISM band, the details and specifications are not open source and can only be accessed by entities who have signed an agreement with Sigma Systems~\cite{fouladi2013security}. This protocol is geared towards medium-short range IoT networks where low power consumption for the nodes is a prerequisite, along with a low data rate. Mesh topology can be implemented for Z-Wave~\cite{al2017internet, salman2015internet, ab2015comparison}.

%%%%%%%%%%%%%%%%%%%%%%%%%%%%%%%%%%%%%%%%%%

\section{Performance comparison}\label{Performance comparison}

\label{section_performance}

This section provides an empirical guideline for selecting the appropriate communication protocol by comparing the IoT communication protocols. In order to compare them, parameters such as standard, energy consumption, coverage, data rate, security, modulation type, cost, and other factors are featured.

\subsection{Data rate}

Among the protocols, LTE-M provides the most significant data rate of up to 1 Mbps, while Sigfox, LoRa, Z-Wave, and NB-IoT have data rates of less than 1 Mbps~\cite{mekki2019comparative, hassan2020nb, dragomir2016survey}. Although Sigfox utilizes the UNB very efficiently and provides high receiver sensitivity and low-cost antenna due to very low-level noise, it falls short of the data rate, offering only 0.1kHz~\cite{mekki2019comparative}. In the case of NB-IoT, compared to the downlink rate, the uplink data rate is lower and upto 20 kbps~\cite{mekki2019comparative}.

\subsection{Security}
All of the IoT communication protocols utilize authentication and encryption procedures in terms of security. Whereas Sigfox employs low-level authentication and encryption, LoRa, Z-Wave, NB-IoT, and LTE-M use the Advanced Encryption Standard (AES) block cipher with counter mode. AES is safer than authentication and encryption. Compared to AES, authentication and encryption is extremely quick but exposed~\cite{chacko2018security, oliveira2019mac}.

\subsection{Range}
NB-IoT and LTE-M have far less coverage compared to the ISM band protocols, with Sigfox having a maximum range of up to 40km, LoRa having a good range of up to 20km, and Z-Wave upto 30km~\cite{mekki2019comparative, hassan2020nb, dragomir2016survey}.
To put the matter into perspective, by using Sigfox, we can cover a typical large city with just one single base station. LoRa has a lower reach, which can take up to three base stations for full coverage of the entire city. NB-IoT and LTE-M primarily focus on applications where the end devices have issues with typical cellular networks or ISM band-based technologies due to physical barriers (indoor applications). In contrast, Z-wave focuses on even shorter-range indoor applications compared to NB-IoT and LTE-M~\cite{mekki2019comparative}.
 
\subsection{Energy Consumption}
LoRa, Sigfox, and Z-Wave are developed for portable devices with low battery capacity because of their low energy usage. They have minimal power usage as a result. In contrast, NB-IoT and LTE-M consume more energy when compared to LoRa and Sigfox. Among these five protocols, LoRa has the highest energy efficiency of them all~\cite{hassan2020nb, mehboob2016survey, aragues2012trends}. 

Although LTE-M and NB-IoT both have medium energy consumption rates, LTE-M performs better in favorable coverage conditions and gets better as the conditions improve. In contrast, NB-IoT performs better in unfavorable coverage conditions~\cite{sorensen2022modeling}. However, for synchronous communication and QoS management, NB-IoT UE devices require more peak current, and its Orthogonal Frequency-division Multiplexing (OFDM)/Frequency Division Multiple Access (FDMA) access modes use more power~\cite{oh2016efficient}.

\begin{table*}[]
\centering
%\begin{adjustwidth}{-2cm}{-2cm}
\scriptsize
\caption{IoT protocols parameters comparative analysis}

\begin{tabular}{cccccc}
\hline
\textbf{Features} &
  \textbf{LoRa}&
  \textbf{Sigfox} &
  \textbf{NB-IoT} &
  \textbf{LTE-M} &
  \textbf{Z-Wave} \\ \hline
Standard &
  LoRaWAN{~\cite{wang2017primer}} &
  \begin{tabular}[c]{@{}c@{}}Collaboration\\ with ETSI{~\cite{lauridsen2018empirical}}\end{tabular} &
  3GPP{~\cite{Fattah2019-ls}} &
  3GPP{~\cite{ratasuk2017lte}} &
  Sigma Designs{~\cite{dragomir2016survey}} \\\\ %\hline
\begin{tabular}[c]{@{}c@{}}Frequency\\ Band\end{tabular} &
  \begin{tabular}[c]{@{}c@{}}Sub GHz \\ ISM Bands{~\cite{wang2017primer}}\end{tabular} &
  \begin{tabular}[c]{@{}c@{}}Sub GHz \\ ISM Bands{~\cite{wang2017primer}}\end{tabular} &
  \begin{tabular}[c]{@{}c@{}}Licensed \\ Bands{~\cite{Fattah2019-ls}}\end{tabular} &
  \begin{tabular}[c]{@{}c@{}}Licensed   \\ Bands{~\cite{borkar2020long}}\end{tabular} &
  \begin{tabular}[c]{@{}c@{}}ISM \\ Bands{~\cite{fouladi2013security}}\end{tabular} \\\\ %\hline
\begin{tabular}[c]{@{}c@{}}Minimum \\ Carrier Bandwidth\\(kHz)\end{tabular} &
  125{~\cite{noreen2017study}} &
  0.1-0.6~\cite{sangar2020wichronos} &
  200{~\cite{hassan2020nb}} &
  1400{~\cite{borkar2020long}} &
  868.42(EU) and 908(US)~\cite{oliveira2019mac, gomez2010wireless} \\\\ %\hline
Data Rate (kbps) &
  50{~\cite{mekki2019comparative}} &
  0.1{~\cite{mekki2019comparative}} &
  250{~\cite{hassan2020nb}} &
  Upto 1000{~\cite{dawaliby2016depth}} &
  Upto 100{~\cite{dragomir2016survey}} \\\\ %\hline
\begin{tabular}[c]{@{}c@{}}Transmission \\ Range (km)\end{tabular} &
  5-20{~\cite{mekki2019comparative}} &
  10-40{~\cite{mekki2019comparative}} &
  1-10{~\cite{mekki2019comparative}} &
  5{~\cite{hassan2020nb}} &
  0.003{~\cite{dragomir2016survey}} \\\\ %\hline
\begin{tabular}[c]{@{}c@{}}Energy \\ Consumption\end{tabular} &
  Very low{~\cite{allLora15}} &
  Low{~\cite{oliveira2019mac}} &
  Medium{~\cite{sinha2017survey}} &
  Medium{~\cite{borkar2020long}} &
  Low{~\cite{aragues2012trends}} \\\\ %\hline
Cost  &
  Low{~\cite{mekki2018overview}} &
  Low{~\cite{mekki2018overview}} &
  Medium{~\cite{mekki2018overview, borkar2020long}} &
  High{~\cite{mekki2018overview, borkar2020long}} &
  Low{~\cite{al2017internet}} \\\\ %\hline
Security &
  AES128{~\cite{tsai2018aes,thaenkaew2022evaluating}} &
  \begin{tabular}[c]{@{}c@{}}Authentication \\ and Encryption{~\cite{chacko2018security}}\end{tabular} &
  AES 256{~\cite{oliveira2019mac}} &
  AES 256{~\cite{oliveira2019mac}} &
  \begin{tabular}[c]{@{}c@{}}AES-128 \\ and CCM{~\cite{dragomir2016survey}}\end{tabular} \\\\ %\hline
Modulation &
  CSS{~\cite{chaudhari2020lpwan}} &
  \begin{tabular}[c]{@{}c@{}}DBPSK, GFSK{~\cite{chaudhari2020lpwan}}\end{tabular} &
  QPSK{~\cite{sinha2017survey}} &
  \begin{tabular}[c]{@{}c@{}}QPSK, 16-QAM \\ and 64-QAM{~\cite{borkar2020long}}\end{tabular} &
  BFSK{~\cite{gomez2012overview}} \\\\ %\hline
\begin{tabular}[c]{@{}c@{}}Battery life time\\ (Years)\end{tabular} &
  \textgreater{}10{~\cite{mehboob2016survey}} &
  \textgreater{}10{~\cite{mehboob2016survey}} &
  \textgreater{}10{{~\cite{hassan2020nb}}} &
  10{{~\cite{hassan2020nb}}} &
  \textgreater{}10{~\cite{hasanaj2019air}} \\\\ %\hline
\begin{tabular}[c]{@{}c@{}}Link budget\\ (db)\end{tabular} &
  154{~\cite{adelantado2017understanding}} &
  159{~\cite{mekki2018overview}} &
  151{{~\cite{allLora15}}} &
  146{{~\cite{allLora15}}} &
  101{~\cite{leussink2012wireless}} \\ \hline
\end{tabular}
\label{table:IoT_Protocols_Parameters_Comparative_Analys}
%\end{adjustwidth}
\end{table*}

\subsection{Cost}
In terms of price, the ISM-band based Sigfox, LoRa, and Z-Wave are all cheaper than the cellular-band based NB-IoT and LTE-M. Although Sigfox and LoRa are significantly cheaper than LTE-M and NB-IoT, respectively~\cite{al2017internet, oliveira2019mac, lauridsen2018empirical, borkar2020long, allLora15}, in between LoRa and Sigfox, LoRa is the more expensive one. LoRa has a long battery life by the framework, so the cost of replacing devices can be reduced. As architecture follows traditional wireless protocol regulation, deployment is straightforward. Similarly, Sigfox is designed to maximize energy efficiency by the ingrained framework. When it transmits data, Sigfox consumes deficient power, and hardly maintenance is required~\cite{oliveira2019mac}. The above-stated features in the design help reduce the cost of LoRa and Sigfox. On the other hand, LTE-M uses power consumption features to extend device health for a long time, although licensed band standard and long coverage framework are more expensive than LPWAN protocol~\cite{al2017internet, oliveira2019mac}. 

\subsection{QoS}

According to studies, cellular NB-IoT and LTE-M offer very low latency~\cite{raza2017low, wang2016modeling}, while protocols like LoRa and Sigfox can offer even lower latency~\cite{mekki2018overview, ferre2018introduction}. LoRaWAN, in contrast to Sigfox, offers lower bidirectional latency.  
Speed and low latency are LTE-M’s significant benefits. LTE-M can deliver speeds of up to 1 Mbps in the uplink and 384 Kbps in the downlink. In addition, LTE-M has a 50–100 ms latency, its nodes can transmit at more effective data rates, and the reduced latency allows for real-time communication between the nodes~\cite{borkar2020long}.\hfill \break

While comparing IoT communication protocols, Sigfox emerged as the future protocol because of its low cost and broad reach. However, a low-power consumption module will work best for LoRa. Overall, LoRa is more compatible with all environments thanks to its security features and low cost, great range, and low power consumption~\cite{mekki2018overview, borkar2020long}. Table \ref{table:IoT_Protocols_Parameters_Comparative_Analys} provides a side-by-side juxtaposition of five dominant IoT communication protocols that are two wide-area protocols based on ISM bands (LoRa and Sigfox), two cellular protocols (NB-IoT and LTE-m), and a short-range ISM-based protocol (Z-wave). 

\begin{figure*}[htbp]
\centerline{\includegraphics[scale=.7
]{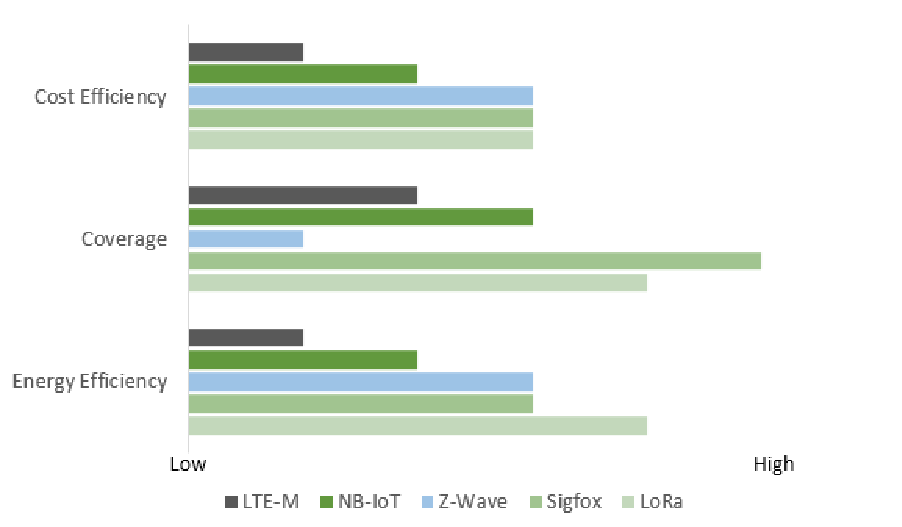}}
\caption{Comparison from cost efficiency, coverage and energy efficiency points of view}
\label{fig_4}
\end{figure*}

Among all the requirements regarding IIoT communication systems, cost efficiency, coverage, and energy efficiency are three features of prime importance. Buying and deployment of IoT devices need to be as inexpensive as possible because a large number of devices are needed to be deployed to create an efficient sensing or monitoring system in the industrial environment. Also, IoT systems will often be used across small to large geo-locations depending on the industry size and purpose that need to be covered by the system. That is why transmission range and coverage are very important factors when deciding the suitable protocol for a scenario. Lastly, in some cases, IoT devices work in such industrial systems where continuous transmission and reception of data is not always necessary, but longevity is essential. So, energy efficiency carries great significance in this area of communication protocols. Figure \ref{fig_4} provides a graphical comparison among the five IoT communication protocols (LoRa, Sigfox, Z-Wave, NB-IoT, and LTE-M) from cost efficiency, coverage, and energy efficiency points of view. The figure shows that LPWAN-based IoT protocols (for example, LoRa, Sigfox, NB-IoT, and LTE-M) are more suitable for industrial applications. Their low power consumption, high coverage, and cost-effectiveness make them indispensable solutions for IIoT applications in diverse use cases. The rest of this paper explores the industry-specific understanding of LPWAN-based IoT deployment in various industrial application areas.

%%%%%%%%%%%%%%%%%%%%%%%%%%%%%%%%%%%%%%%%%%

\section{Industrial Application Perspective}\label{section_perspective}

\begin{figure*}[htbp]
\centerline{\includegraphics[width=.75\linewidth
]{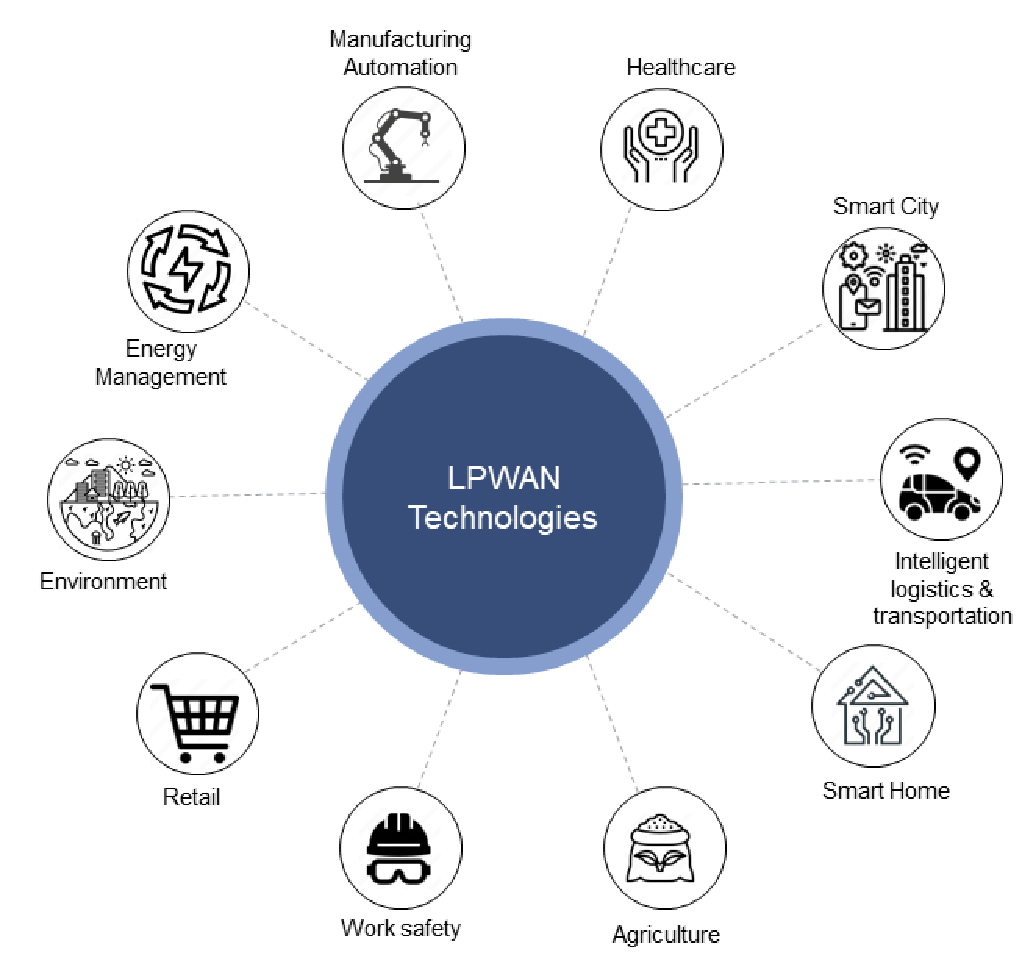}}
\caption{Applications of Industrial Internet of things based on LPWAN protocols}
\label{fig_5}
\end{figure*}

In order to evaluate the applicability of LPWAN and cellular IoT protocols and comprehend the benefits and constraints of the technology, a variety of LPWAN based industrial applications are taken into account. Industrial Internet of things (IIoT) applications have a comprehensive area of use. This section discusses several industrial application use cases and provides an overview of the technologies that best fit each situation in terms of characteristics. Figure \ref{fig_5} illustrates the prime application area of LPWAN-based protocols in IIoT.

\subsection{Smart City} 

Large-scale uses of IoT technology, such as intelligent parking, automated lighting, and smart garbage collection, are being developed~\cite{adelantado2017understanding, martinez2015lean} for Smart City applications. Similar to smart garbage collection systems, smart lighting systems respond to a measure with long variable periods by acting or reporting information~\cite{nor2017smart}. Even though there is no significant dependency upon latency and jitter, in certain cases, the triggering factor is concurrent for many UE devices. LoRaWAN and Sigfox are acceptable solutions in this kind of scenario since they can manage the large coverage area and a sizable number of users at the cost of increasing collision, latency, and jitter rates~\cite{al2017internet, adelantado2017understanding, allLora15}.

\subsection{Intelligent Logistics and Transportation} 

Logistics and transportation are two essential foundations of the anticipated IoT development over the coming years. Most applications aim to increase efficiency in sectors like cargo or public transportation. However, although specific applications can tolerate jitter, delay, or unreliability, others cannot. Therefore, based on the IEEE 802.11p standard, other standards have been created in the 5.9 GHz band for Intelligent Transportation Systems (ITS)~\cite{shahgholi2021lpwan}. 

Due to the enormous quantity of data generated by sensors deployed on cars or Road Side Units (RSUs), ITS systems may have extra communication overhead, high bandwidth consumption, and more significant reaction delays while transmitting data to cloud servers. The cellular LPWAN architectures LTE-M and NB-IoT have been used by ITS applications to meet the challenges mentioned earlier. Also, LTE-M and NB-IoT are appropriate for providing backhaul infrastructure for ITS applications~\cite{shahgholi2021lpwan}. NB-IoT and LTE-M are the best suited for these applications because of their diversity, range, QoS, and low latency~\cite{shahgholi2021lpwan}. However, LoRaWAN, an LPWAN solution, is inappropriate for these applications~\cite{adelantado2017understanding}.

\subsection{Smart Farming and Agriculture} 

Long battery lives are needed for sensor equipment in the agricultural sector. In a conventional farming environment, short coverage is sufficient. According to~\cite{mekki2018overview, ray2017internet}, utilizing temperature and humidity sensors in the irrigation industry might drastically cut water use while increasing productivity. Considering a farming environment generally remains the same, devices only update sensor data a few times per hour. Sigfox and LoRaWAN are perfect for this sort of technology due to their low power consumption~\cite{mekki2018overview}. Also, many agricultural businesses still need LTE-cellphone based application connectivity due to financial and technological constraints. As a result, the NB-IoT cloud might not be a viable solution for agriculture in the future.

\subsection{Smart Home}

Property supervisors often use alerting measures such as temperature, humidity, safety, water flow, and electric plug sensors to prevent damage and promptly respond to requests without monitoring. This is a normal use of a clever home framework. According to~\cite{mekki2018overview, nanni2017indoor}, these sensors need reasonably priced, long-lasting batteries. Sigfox, LoRaWAN are better suited for this type of application since they do not require frequent communication or high-quality service, and short-range operation is acceptable.

\begin{table*}[]
\centering
%\begin{adjustwidth}{-\extralength}{0cm}
\small
\caption{Suitable protocol based on application and physical features}
\begin{tabular}{ccc}
\hline
\multicolumn{1}{c}{\textbf{Use Cases}} & \multicolumn{1}{c}{\textbf{Features}}                                                              & \multicolumn{1}{c}{\textbf{Protocol}}    
                                                     \\ \hline\\
Smart City                               & Large Coverage Area, Latency, Cost                                                                  & Sigfox and LoRaWAN                                                       \\\\
Intelligent Logistics and Transportation & Low latency, High QoS, Large Coverage                                                               & NB-IoT                                                                   \\\\
Smart Farming and Agriculture            & Large Coverage Area, Latency, Cost, Low Power                                                       & Sigfox and LoRaWAN                                                       \\\\
Smart Home                               & Short range, Lower Latency, Cost, Low Power                                                         & Sigfox and LoRaWAN                                                      \\\\
Terminals for Retail Sales               & Low Latency, High QoS, Large Coverage                                                               & NB-IoT                                                                   \\\\
Smart Environment                        & Low Latency, High QoS, Large Coverage                                                               & NB-IoT, LTE-M                                                            \\\\
Smart Metering, Energy, and Grid         & \begin{tabular}[c]{@{}c@{}}Long Range, Low Power, Robust \\ QoS, and Readability\end{tabular}       & NB-IoT                                                                   \\\\
Manufacturing \& Automated Industries    & \begin{tabular}[c]{@{}c@{}}Long Range, Low Power, Robust QoS, \\  Readability and Cost\end{tabular} & \begin{tabular}[c]{@{}c@{}}Sigfox and LoRaWAN \\ and NB-IoT\end{tabular} \\\\
Wearables \& Health                      & \begin{tabular}[c]{@{}c@{}}Long Range, Low Power, Robust QoS, \\  Readability and Cost\end{tabular} & \begin{tabular}[c]{@{}c@{}}Sigfox and LoRaWAN \\ and NB-IoT\end{tabular} \\\\
Work Safety                              & Low Power, Cost and Readability                                                                     & \begin{tabular}[c]{@{}c@{}}Sigfox and LoRaWAN \\ and NB-IoT\end{tabular} \\ \hline
\end{tabular}
%\end{adjustwidth}
\end{table*}

\subsection{Terminals for Retail Sales}

Since they deal with regular contact, sale point systems demand guaranteed quality of service~\cite{mekki2018overview, saarikko2017internet}. There is no limit on battery life because these devices have a constant electrical power source. Low latency is also crucial; otherwise, it limits the transactions a store can process within a given time~\cite{mekki2018overview}. As a result, NB-IoT and LTE-M are more appropriate for this application. Cost is also a consideration for retail point-of-sale terminals, making NB-IoT preferable over LTE-M.

\subsection{Smart Environment}

IoT-based innovative environments contain information about water quality, lowering levels of pollution in the air, lowering temperatures, preventing forest fires and landslides, tracking animals, monitoring snow levels, and early detection of earthquakes~\cite{chaudhari2020lpwan}. This type of project calls for sensors with long battery lives and also takes coverage and range into account. However, they also require high QoS, large bandwidth, and efficient bypassing of interference. Also, the projects are often carried out by large-scale undertakings that can make higher expenses bearable. As a result, NB-IoT and LTE-M are more appropriate for this sort of application. However, the administration of smart water grids can benefit from the deployment of LoRaWAN~\cite{saravanan2017smart}.

%\subsection{Smart Metering, Energy, and Grid} 
\subsection{Energy Management} 

In order to build an industrial-level smart grid energy metering environment based on IoT, network control, load adjusting, remote observing and estimation, transformer wellbeing checking, and observation of wind plants/sunlight-based power establishments are a few significant variables~\cite{al2017internet, chaudhari2020lpwan}. Long-range, low power, robust QoS, and excellent readability are requirements for high-level smart grid and energy metering. NB-IoT is, hence, better suited for usage in this application. Additionally, at the point where the power distribution section consists of individual commercial and residential appliances, LoRaWAN technology simplifies the transition to the Private Area Network (PAN) and Home Area Network (HAN)~\cite{jain2018smart}. This includes smart electric bill meters, smart hazard alarm systems, etc. 

%\subsection{Energy Management} 

The energy sector has been altered by LPWAN technology, and efficient sensor monitoring systems have reduced factory energy use~\cite{judge2021secure}. As a result, the industrial energy system is a critical component of the IIoT. Furthermore, IIoT technologies have improved the efficiency of modern energy systems. For example, advanced control systems, predictive maintenance, and remote monitoring can improve smart energy management~\cite{farooq2023survey}. Big data analytics, software-defined machines, and smart sensors are emerging technologies that have steadily improved the system's operating performance~\cite{wan2016software}.

\subsection{Manufacturing and Automated Industries} 

LPWAN technologies play a vital role in developing cost-effective solutions in predictive industrial maintenance. Predictive maintenance involves the monitoring of the industrial equipment \cite{sanchez2021design} and predicting the required maintenance \cite{pointl2021assessing} when necessary. It can reduce the possibility of unwanted shutdown of the equipment and reduce costs. LPWAN also provides solutions for effective and cost-efficient asset monitoring \cite{chaudhari2020lpwan} due to its long-range communication and low power consumption capabilities. LPWAN enables continuous and real-time tracing of the assets that help in optimizing asset utilization and improving overall resource allocation \cite{kaburaki2023adaptive}. It also improves the supply chain visibility \cite{khan2023applications} by tracing and monitoring the shipment of goods and optimizing the supply chain operations.

% There are many different kinds of sensors and communication requisites in factory automation. Sigfox and LoRaWAN are appropriate alternatives to NB-IoT for specific applications, such as asset tracking and status monitoring~\cite{chaudhari2020lpwan}. However, for other applications, like status monitoring, low-cost sensors with long battery lives are required. Sigfox and LoRaWAN are better options in these situations. Due to the wide range of requirements, hybrid solutions may also be utilized. 

% In \cite{paolini2021rf}, the design of an LPWAN-based self-deployable wireless RF system is proposed for remotely monitoring Electromagnetic extreme environments.

\subsection{Wearables and Health}

Using different IoT communication protocols, it is simple to observe and work on a patient’s health-related parameters, operate connected medical environments, healthcare wearables, patient surveillance, telemedicine, fall detection, athletes care, track chronic diseases and track mosquito and other similar insect populations~\cite{chaudhari2020lpwan}. Most systems require low latency, diversity, range, and QoS, making NB-IoT a better fit for these applications. 

The healthcare business is tremendously benefiting from IIoT applications. They save money by allowing remote control of medical equipment, home-bound patient care, modelling, and monitoring~\cite{al2018context}. As a result, hospitals benefit from innovative equipment that reduces patient wait times and enhances equipment performance. The growing popularity of mobile internet connections has accelerated the expansion of IIoT-powered in-home healthcare (IHH) services~\cite{farooq2023survey}.
Patient-Centered Medical Home (PCMH) care is a straightforward answer to numerous problems in the healthcare sector, such as chronic disease management, misuse of emergency departments, patient satisfaction, excessive medical expenses, and accessibility. Sensor devices provide significant information about patient health and aid in diagnosing disease~\cite{kumar2018cloud}. Furthermore, the IIoT application domain provides telemedicine solutions, such as notifying patients about their well-being and monitoring their health with advanced medical equipment~\cite{aceto2020industry}.

Healthcare wearables, patient monitoring, and indoor remote health, PCMH care and wellbeing monitoring are examples of applications that can be operated with shorter-range IoT protocols like Sigfox, LoRa~\cite{iqbal2020application, petajajarvi2017evaluation, chung2018experiments}.

\subsection{Work Safety} 
Enhanced security for IIoT risk management and control security principles, many security policies exist. Furthermore, the IIoT energy system can detect defects and energy usage of different components by continuous monitoring and real-time data processing. As a result, the system can prevent severe and harmful accidents and wasteful losses while also increasing overall energy efficiency. Monitoring and response systems can be well suited with long-range and low-power IoT protocols like LPWAN~\cite{mouratidis2018security}. 

Table \ref{table:Protocol based on Application} summarizes the relevancy of the different IoT protocols from an application perspective to point out which protocol is suitable for certain usage criteria.

%%%%%%%%%%%%%%%%%%%%%%%%%%%%%%%%%%%%%%%%%%

\section{Open Issue}\label{section_future}

Although the aforementioned critical technologies lay the groundwork for future IoT connectivity, they only serve to demonstrate the qualitative viability of the aim. Moreover, many problems still need to be considered as IoT is a relatively young technology. In this section, we examine some of these issues for future investigation. 

\subsection{Combining Multiple Protocols}
 
Integrating different protocols in the standard IoT architecture across different areas of the IoT cloud is challenging. But this diversity is necessary for developing concepts for the future internet. However, it is not possible to utilize a combined protocol architecture if one protocol uses a different architecture than the other~\cite{dizdarevic2019survey}.

\subsection{Security} 
The topic of privacy is still poorly covered in most research, including anonymous communication and filtering system issues. This field still needs significant work in order to improve, even though several articles address essential IoT or particular protocol privacy challenges~\cite{frustaci2017evaluating}.

\subsection{Single Point Gateway} 

Data is sent from sensors to LPWAN gateways. The gateways establish an IP connection to the Internet and send the data obtained from the embedded sensors to the Internet, which can be a network, a server, or a cloud. The gateways function as a transparent bridge, translating Radio Frequency (RF) transmissions to IP packets and vice versa, connecting to the network server via conventional IP connections. However, due to LPWAN technology’s single point of failure at the gateway section and a lack of redundancy, using gateways to communicate with end devices may become inefficient~\cite{mehboob2016survey}.

\subsection{ALOHA-based Access}

Advocates of Linux Open-source Hawaii Association-based (ALOHA) access could be optimized for serving deterministic traffic, which is acquiring significance in the IoT ecosystem. Sadly, deterministic traffic handling is a typical restriction of the IoT mac layer protocol~\cite{adelantado2017understanding}, and this needs to be addressed. 
 
\subsection{Data Management}

Data extraction, which can be viewed as gathering data from the appliances and extracting meaningful information from the obtained data, is an open issue that needs to be considered. Data extraction will significantly affect how well a system functions, notably when the number of appliances is expanded in a communication architecture. If the system has to be completely redesigned, it may be determined using the memory capacity, processing speed, and network bandwidth. In contrast to the problems with data extraction, data representation is a crucial area for research because it may facilitate information interchange between the IoT communication system and other technologies like ontology and semantic web technologies~\cite{tsai2014future, katasonov2008smart}.\hfill \break

Apart from the above-mentioned challenges, some other facts about IoT communication protocols must also be addressed. Numerous analysis studies of the protocols may consider the following topics, estimation of the collision rate, channel load, single device maximal throughput and Maximum Transmission Unit (MTU), mobility/roaming, and proposing possible solutions for performance enhancement~\cite{de2017lorawan}.

%%%%%%%%%%%%%%%%%%%%%%%%%%%%%%%%%%%%%%%%%%
\section{Conclusion}\label{section_conclusion}

Due to scalability being one of the most important factors in its practical applications, IoT devices will be deployed in large numbers within a few years. However, so many architectures and protocols exist among IoT devices that it can easily become confusing which one is more suitable in an industrial environment and which one to avoid there. Some of the major technologies in the IoT market are Sigfox, LoRa, Z-Wave, NB-IoT, and LTE-M. We present an overview of the basic network architectures of these IoT communication protocols and a comparison that focuses on some of the key factors, for example, low cost, long-range, and energy efficiency. %of the protocols as they play vital roles  %in choosing the right kind of device in an environment covering a large physical area, longer runtime thanks to lesser power consumption, and all of these at a lower expense. So far, we have found that Sigfox has the lowest cost while LoRa offers the lowest power consumption. %And both of these work free of costISM band.
The results of performance comparison and application perspective show that LPWAN-based protocols perform better and are more suitable than other IoT Protocols. Several parameters are essential to deciding which technology should be used in industrial environments, which is covered in this study from an application perspective. %The main purpose of this survey is to provide a clear comparison among the major architectures in the IoT industry and be a reference in decision-making.
However, there are some challenges in this relatively novel field of research that have also been addressed in this paper. 

%ISM-based Z-Wave can support IoT services where short-range communication is sufficient to handle the tasks, while LoRa and Sigfox can be used for long coverage necessities. On the other hand, 3GPP-based NB-IoT and LTE-M offer comparatively medium ranges, although they consume more power at higher costs. Still, the 3GPP-based technologies work in the licensed bandwidth spectrum and can offer a much better QoS and robustness compared to the ISM users. 

%%%%%%%%%%%%%%%%%%%%%%%%%%%%%%%%%%%%%%%%%%
% \section{Patents}

% This section is not mandatory, but may be added if there are patents resulting from the work reported in this manuscript.

%%%%%%%%%%%%%%%%%%%%%%%%%%%%%%%%%%%%%%%%%%
\vspace{6pt}

\section*{Abbreviations}{
The following abbreviations are used in this manuscript:\\

\noindent 
\begin{tabular}{@{}ll}
3GPP & 3rd Generation Partnership Project \\
AES	& Advanced Encryption Standard \\
AMQP & Advanced Message Queuing Protocol \\
ALOHA & Advocates of Linux Open-source Hawaii Association-based \\
CSS & Chirp Spread Spectrum \\
CoAP & Constrained Application Protocol \\
EPC &Evolved Packet Core \\
FDMA & Frequency Division Multiple Access \\
HAN	& Home Area Network \\
HSS	& Home Subscriber Server \\
ISM	& Industrial, Scientific and Medical \\
ITS	& Intelligent Transportation Systems \\
LTE-M & Long Term Evolution for Machines \\
LPWAN & Low Power Wide Area Network \\
M2M	& Machine-to-Machine \\
MTU	& Maximum Transmission Unit \\
MME	& Mobility Management Entity \\
MQTT & MQ Telemetry Transport \\
NB-IoT & Narrow Band Internet of Things \\
OFDM & Orthogonal Frequency-division Multiplexing \\
PGW	& Packet Gateway \\
PAN	& Private Area Network \\
QPSK & Quadrature Phase Shift Keying \\
% QoS	& Quality of Service \\
% RF & Radio Frequency \\
RSUs & Road Side Units \\
SGW	& Serving Gateway \\
UNB	& Ultra Narrow Band \\
% WAN	& Wide Area Networks \\
WSN	& Wireless Sensor Networks \\

\end{tabular}
}

%%%%%%%%%%%%%%%%%%%%%%%%%%%%%%%%%%%%%%%%%%
%% Optional
% \appendixtitles{no} % Leave argument "no" if all appendix headings stay EMPTY (then no dot is printed after "Appendix A"). If the appendix sections contain a heading then change the argument to "yes".
% \appendixstart
% \appendix
% \section[\appendixname~\thesection]{}
% \subsection[\appendixname~\thesubsection]{}
% The appendix is an optional section that can contain details and data supplemental to the main text---for example, explanations of experimental details that would disrupt the flow of the main text but nonetheless remain crucial to understanding and reproducing the research shown; figures of replicates for experiments of which representative data are shown in the main text can be added here if brief, or as Supplementary Data. Mathematical proofs of results not central to the paper can be added as an appendix.

% \begin{table}[H] 
% \caption{This is a table caption.\label{tab5}}
% \newcolumntype{C}{>{\centering\arraybackslash}X}
% \begin{tabularx}{\textwidth}{CCC}
% \toprule
% \textbf{Title 1}	& \textbf{Title 2}	& \textbf{Title 3}\\
% \midrule
% Entry 1		& Data			& Data\\
% Entry 2		& Data			& Data\\
% \bottomrule
% \end{tabularx}
% \end{table}

% \section[\appendixname~\thesection]{}
% All appendix sections must be cited in the main text. In the appendices, Figures, Tables, etc. should be labeled, starting with ``A''---e.g., Figure A1, Figure A2, etc.

% %%%%%%%%%%%%%%%%%%%%%%%%%%%%%%%%%%%%%%%%%%
% \begin{adjustwidth}{-\extralength}{0cm}
% %\printendnotes[custom] % Un-comment to print a list of endnotes

\bibliographystyle{unsrt}  
\bibliography{references}

\end{document}